\documentclass[12pt]{article}
\usepackage{amsmath}
\usepackage{amssymb}
\usepackage{tabularx}
\usepackage{enumerate}

\usepackage{mathrsfs}
\usepackage{amssymb, upref, color}
\usepackage[dvips]{graphicx}
\usepackage{threeparttable}

\usepackage{texdraw}

\newtheorem{thm}{Theorem}[section]

\newtheorem{defi}{Definition}[section]
\newtheorem{pppp}{Proof}

\newcommand{\qed}{\hspace{1em}\mbox{\raisebox{0.65ex}{\fbox{}}}}

\numberwithin{equation}{section}

\newcommand{\be}{\begin{equation}}
\newcommand{\ee}{\end{equation}}
\newcommand\bes{\begin{eqnarray}}
\newcommand\ees{\end{eqnarray}}
\newcommand{\bess}{\begin{eqnarray*}}
\newcommand{\eess}{\end{eqnarray*}}

\begin{document}
\setlength{\baselineskip}{16pt} \pagestyle{myheadings}

\title{Four-tier response system and spatial propagation of COVID-19 in China
by a network model
\thanks{The work is partially supported:  Lin by the NNSF of China (11771381); Ge by the NNSF of China (11701206)
and Zhu by CIHR and Canadian COVID-19 Math Modelling Task Force .}}
\date{\empty}
\author{Jing Ge$^1$, Daihai He$^2$, Zhigui Lin$^3$\thanks{Corresponding author. Email: zglin68@hotmail.com(Z. Lin).}, Huaiping Zhu$^4$ and Zian Zhuang$^2$\\
{\small 1 School of Mathematics and Statistics, Huaiyin Normal University, Huaian 223300, China}\\
{\small 2 Department of Applied Mathematics, Hong Kong Polytechnic University, Hong Kong, China}\\
{\small 3  School of Mathematical Science, Yangzhou University, Yangzhou, 225002, China}\\
{\small 4 Lamps and Centre for Disease Modelling (CDM), Department of Mathematics and Statistics,}\\
{\small York University, Toronto, ON, M3J 1P3, Canada}
}
\maketitle

\begin{quote}
\noindent {\bf Abstract.} {\small
In order to investigate the effectiveness of lockdown and social distancing restrictions, which have been widely carried out as policy choice
to curb the ongoing COVID-19 pandemic around the world, we formulate and discuss a staged and weighed networked system based on a classical SEAIR epidemiological model.
 Five stages have been taken into consideration according to four-tier response to Public Health Crisis, which comes from the National Contingency Plan in China.
Staggered basic reproduction number has been derived and we evaluate the effectiveness of lockdown and social distancing policies under
different scenarios among 19 cities/regions in mainland China.
Further, we estimate the infection risk associated with the sequential release based on population mobility between cities and
the intensity of some non-pharmaceutical interventions.
Our results reveal that Level I public health emergency response is necessary for high-risk cities, which can flatten the COVID-19 curve effectively and quickly. Moreover,
 properly designed staggered-release policies are extremely significant for the prevention and control of COVID-19,
furthermore, beneficial to economic activities and social stability and development.}

  \vspace{0.5cm}

%----------------------------------------------------------------------
\end{quote}

\newcommand\HI{{\bf I}}

\section{Introduction}

 Since late December of 2019, an increasing number of atypical pneumonia cases caused by severe acute respiratory syndrome coronavirus (SARS-CoV-2)
 have been reported in Wuhan, China \cite{LKK}. On January 30, 2020, the World Health Organization (WHO) declared COVID-19 outbreak a
 Public Health Emergency of International Concern (PHEI).
 However, the COVID-19 has been kept spreading around the world at an alarming speed, and has now spread to more than 200 countries, areas, or territories. On March 11, 2020, WHO declared the spread of COVID-19 as a global pandemic \cite{WHO}. As of August 14, 2020, there have been more than 21 million confirmed cases of COVID-19 worldwide with more than 757,000 deaths \cite{WHO}.

To fight the highly contagious new coronavirus that has ravaged the world in such a short period of time, most of the countries around the world have taken strict prevention and control measures to mitigate and curb the spreading of the pandemic. In the absence of effective antiviral treatment or vaccines, it is now popular to implement non-pharmaceutical interventions (NPIs), including
 social distancing, contact tracing and testing, self-quarantine or isolation,
 closing schools and workplaces, banning the gatherings. In particularly, in order to block the spreading, more than 100 countries carry out full or partial lockdown strategy to battle the pandemic \cite{BHR, BD, EEL, FNP}.

Among some of the few successful countries, China was the one to have stopped the spatial propagation of the virus earlier \cite{Krae},
and now the country is gradually coming back to normal.
There have been extensive modeling studies about the transmission of the virus and how Wuhan and other provinces in China had successfully controlled the epidemic. Recall that after the SARS outbreak in 2003,
  China has established a legal framework with the Law on the Prevention and Treatment of Infectious Diseases as its core part.
In response to public health emergencies, the National Contingency Plan for Public Health Crisis is formulated, and
the Plan divides public health emergency response into four-tier: extremely significant (Level I or Top level), major (Level II), relatively major (Level III), and general (Level IV), with level I the most serious \cite{NHC}.

 The outbreak of COVID-19 in Wuhan was unprecedent. After it was recognized that the virus was some novel coronavirus
 and can transmit from one individual to another quickly\cite{RAl},
Wuhan quickly adopted the response system and treated as Level I emergence. On January 23, 2020,  the megacity was lockdown.
On January 25, a total of 26 provinces, municipalities, and autonomous regions across China have initiated
Level I public health emergencies response, covering more than 1.2 billion population. Hence, after locking down and with strict NPIs to limit population mobility and to reduce the local community transmission in the whole country, the level of emergency response gradually dropped
from high to low level till the prevalence of COVID-19 was effectively controlled as of the early of April, 2020 \cite{NHC1}.

Naturally, different control policies will result in different control effectiveness on the transmission of infectious diseases.
Assessing the infection risk and compare the effectiveness of NPIs have become essential tools for public health policymakers.
Nowadays, a large number of mathematical models have been developed to incorporate some NPIs to
explain the mechanisms of the propagation of the ongoing COVID-19 pandemic \cite{EEL, LPS, PLC, ZXT}.
Numerous data-driven modeling studies have been developed
play important role to inform the public health and policymakers \cite{ HPS, PSZ, RVH, ZJY}.
Some predictive models were developed to serve as
 early warning tools to estimate the final infection size of the outbreak and spreading speed,
and assess the effectiveness of the NPI control measures.
Especially, most of the above models
are based on SEIR-type (Susceptible-Exposed-Infectious-Removed) compartmental models \cite{ACW1, ACW2, BHR, BD, FNP, LSM, LKK, LPS, PSZ, Sunw},
 and few takes into account the spatial spread
of the COVID-19.

Factually, even with the strictest prevention and control measures for
the lockdown, individual's basic living needs, such as going to the hospital for medical treatment and the basic food supply chain,
cannot be stopped. Population mobility will inevitably happen in intercity and intracity, nonetheless, which play significant roles in shaping the epidemic spreading process.
Therefore, many multi-patch models have been proposed to explore the influence of population mobility on the spatial spread of disease between patches
with different level of disease prevalence \cite{HDW, songlou, Zou}.

On the other hand, population mobility in epidemic models can be characterized via network or graph.
In the past two decades, epidemic transmission in networked models have been extensively explored, see details in \cite{BHR, JS, LS, PCV, TLR}
and the references therein.
Recently, Population mobility in a SIRS epidemic model was considered via graph
Laplacian diffusion \cite{TLR}, the discrete Laplacian operator has been introduced to describe the transmission between each two cities.

Most research in networked epidemic dynamics have
been carried out with the non-weighted graph, i.e., the passenger numbers between each two cities are constant.
However, it turned out more practically that the passenger number between each two cities could always be different.
For controlling the outbreak of COVID-19 in China, public health has taken some prevention and control measures strictly following the guideline of
four-tier response system, and on different level of epidemic, the extent of NPIs were taken accordingly. Then, on each level, the contact transmission probability between susceptible and infectious individuals varies over time.
For example, in different levels of the epidemic, the travel restrictions  are different to maintain the contact transmission probability at a low level.

In present paper, we will develop a novel weighted network epidemic model by accounting for time-varying population mobility in different stages according to the adjustment of the four-tier response system. We will investigate the impacts of transportation restrictions and NPIs taken in different stages on the propagation of the raging COVID-19 around the world.

\section{A weighted network model at different level of response}
\begin{figure}
    \centering
    \includegraphics[width=1\textwidth]{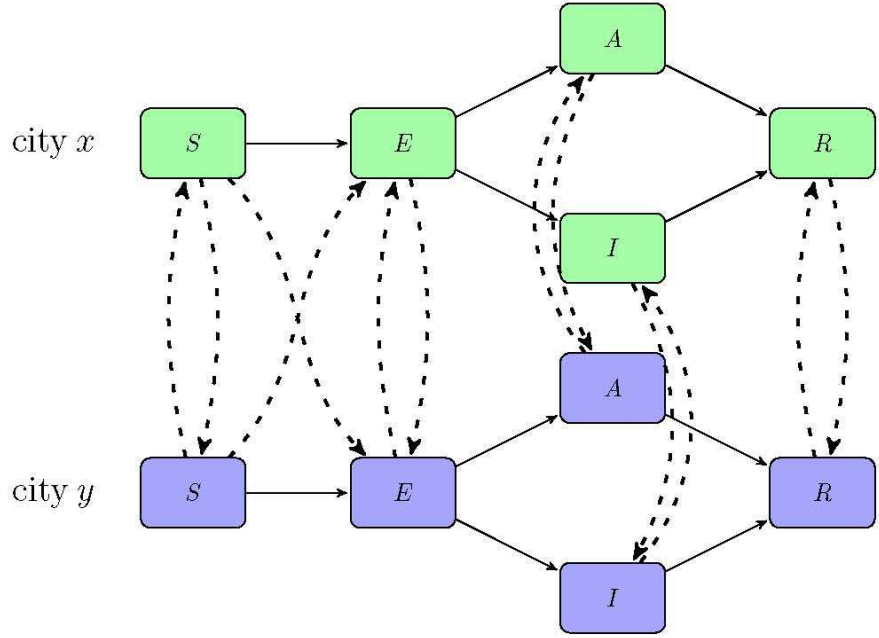}
    \caption{The flowchart of the transmission dynamics model, where the population is stratified by susceptible ($S$),
     exposed ($E$), infectious but asymptomatic ($A$), infectious with symptoms ($I$), and removed ($R$). The solid lines indicate the transmissions inside a city and dashed lines show
     the transmissions between two cities.}
    \label{fig:0l}
\end{figure}

Recently, the  classical SEAIR (Susceptible-Exposed-Asymptomatic-Infectious-Removed) compartmental models
have been used to look into the transmission dynamics of COVID-19.
If we denote susceptible $S$ (which are under risk of
contracting the infection), exposed $E$, infectious but asymptomatic $A$, infectious with symptoms $I$,
 and removed $R$, see Figure \ref{fig:0l}, then a typical ODE model for COVID-19 can be written as
\begin{eqnarray}
\left\{
\begin{array}{lll}
\dfrac{dS}{dt}=-\beta S\frac{I+ \xi A}{N},  \\
\dfrac{dE}{dt}=\beta S\frac{I+\xi A}{N}-\sigma E,\\
\dfrac{dA}{dt}=(1-\theta)\sigma E-\gamma_A A,\\
\dfrac{dI}{dt}=\theta\sigma E -\gamma_I I,\\
\dfrac{dR}{dt}=\gamma_A A+\gamma_I I.
\end{array} \right.\label{A1}
\end{eqnarray}
The $N=S+E+A+I+R$ is the total population. The class $R$ includes the hospitalized population, recovered population and the dead due to disease,
therefore the class is not infectious to others.
The natural death and the disease-related death are ignored since the corresponding parameters (for example, the natural death rate is
about $\frac 1{60\times 365})$ are very small.
All other parameters are summarized in Table 1.
{\small\begin{table}[htbp]
\small
\centering
\caption{\label{table1}Parameters Epidemiological Interpretations. }
\begin{tabular}{|c|c|c| c|c|}
\hline
Parameter      &      Description  & Value &   Sources     \\ \hline

$\beta$     & intracity transmission rate & 0.4 & \cite{EMPE} \\  \hline
$ \xi $    &      the correction factor of transmission probability & 0.5  &  \cite{FN, LPC} \\
& with asymptomatic infectious individuals     & &  \\
   \hline
$ {1}/{\sigma}$    &    the mean incubation period   & 3 & \cite{GNHL}     \\
 \hline
$\theta$    &    the proportion of symptomatic infectious  & 0.7 & \cite{Zhu, Feng} \\
& among the infectious people   & &  \\
\hline

$\gamma_A$    &  rate of asymptomatic infectious individuals   & 0.1 & \cite{EMPE} \\
& losing infectiousness per day      & &  \\
  \hline

$\gamma_I$    &  rate of symptomatic infectious individuals   & 0.25 & \cite{HWLR} \\
& losing infectiousness per day      & &  \\
  \hline
\end{tabular}
\end{table}}

The SEAIR type model is a deterministic meta-population transmission model that simulates each individual in the population as a separate compartment,
and predicts the disease transmission mechanism, which depends on the parameters, the start time and initial values \cite{BN89, HPS, PSZ, ZJY}.
Obviously, system \eqref{A1} neglects the spatial dispersal of the population.
To understand the spatial propagation of the virus, we will employ the discrete Laplacian operators defined on a network to describe the individual's
mobility between cities into the epidemic dynamic system \eqref{A1} \cite{BHL, TLR}.

In what follows, for the readers' convenience, we introduce some definitions and notations about the graph in network structure.
Networks are mathematically characterized as graphs. A
graph $G = G(V, E)$ is a collection of points $V$, which called vertices (nodes in
the physics literature).
These points are joined by a set of connections $E$, which called
edges. Each edge denotes the presence
of a relation or interaction between the vertices it joins. Specifically, $x\in V $ if $x$ is a vertex in $G(V, E)$, some pairs of vertices are connected by links called edges, and $E=\{(x, y)\}$  where $(x, y)$ is an edge between the vertices $x$ and $y$. Two vertices $x$ and $y$ are said to be adjacent if they are connected directly by an edge, and we write $x\backsim y$ to denote that $x$ and $y$ are adjacent.

A weighted graph $G(V, E; \omega)$ is a graph $G(V, E)$ with a weight function
$\omega: V \times V \times [0, \infty) \to [0,+\infty)$ satisfying
$ \omega(x, y; t)\geq 0$ and $ \omega(x, x; t)=0$ for $x, y \in V$,  and
$\omega (x, y; t) > 0$ if and only if $x\backsim y,\, x, y \in V.$
Similarly as stated in \cite{GLIN, PCV, TLR}, we assume that total $n$ cities are in danger of COVID-19 and each city is viewed as a vertex. The function $\omega(x, y; t)$
can be thought of as the probability
per day that an individual of corresponding compartment travels from city $y$ to
city $x$ at time $t$.
\begin{defi}
For the function $f(x,t):\ V\times [0, \infty)\to \mathbb{R}$,
the discrete Laplacian operator $\Delta_\omega$ of $x$ on $V$ at time $t$ is defined by
$$\Delta_\omega f(x,t):=\sum\limits_{y\sim x, \, y\in V}\big(\omega(x,y,t)f(y, t)-\omega(y,x,t)f(x, t)\big)$$
and the discrete gradient operator $\nabla_\omega$ of $x$ on $V$ at time $t$ is defined by
$$\nabla_\omega f(x,t):=\sum\limits_{y\sim x, \, y\in V} \omega(x,y,t)f(y, t).$$

\end{defi}

By regarding graph as a spacial domain, we extend model \eqref{A1}  incorporating the discrete Laplacian operator via finite weighted graph to the model reads as
\begin{eqnarray}
\left\{
\begin{array}{lll}
\frac{\partial S}{\partial t}-\triangle_\omega S+ \beta^*(t)\nabla_\omega (\varepsilon I+ \xi A)=-\beta(t)S\frac{I+ \xi A}{N},&(x,t)\in V\times(0,+\infty), \\
\frac{\partial E}{\partial t}-\triangle_\omega E-\beta^*(t)\nabla_\omega (\varepsilon I+ \xi A)=\beta(t) S\frac{I+ \xi A}{N}-\sigma E,&(x,t)\in V\times(0,+\infty),\\
\frac{\partial A}{\partial t}-\triangle_\omega A=(1-\theta)\sigma E-\gamma_A(t) A,&(x,t)\in V\times(0,+\infty),\\
\frac{\partial I}{\partial t}-\varepsilon\triangle_\omega I=\theta\sigma E -\gamma_I(t) I,&(x,t)\in V\times(0,+\infty),\\
\frac{\partial R}{\partial t}-\triangle_\omega R=\gamma_A(t) A+\gamma_I (t) I,&(x,t)\in V\times(0,+\infty),
\end{array} \right.\label{A3}
\end{eqnarray}
where $\omega(x, y, t)$ describes the population mobility from city $y$ to city $x$,
$\Delta_\omega E(x,t)$ means that $\omega(y, x, t)E(x,t)$ exposed population in city $x$ move to city $y$, at the same day, $\omega(x, y, t)E(y,t)$ exposed population in city $y$ move to city $x$. The coefficient $\varepsilon$ is imposed on the infectious with symptoms based on the assumption that only small portion of infectious with symptoms can travel successfully. Moreover, during the travel from city $y$ to city $x$, the mobile susceptible population $\omega(x, y, t)S(y, t)$ inevitably come into contact with the mobile infectious population $\omega(x, y, t)(I+ \xi A)(y, t)$ , hence, it may have more probability to be exposed with contact transmission rate $\beta^* (:=k\beta >\beta)$.
Therefore the travel leads to an increasing of the exposed in city $x$ with number
$$\sum\limits_{y\sim x,\, y\in V} \beta^*(t)\frac {\omega(x, y, t)S(y, t)}{\omega(x, y, t)N(y, t)} \omega(x,y,t) (\varepsilon I+ \xi A)(y,t)\approx \sum\limits_{y\sim x,\, y\in V} \beta^*(t) \omega(x,y,t) (\varepsilon I+ \xi A)(y, t)$$
 due to all population are almost susceptible. The flowchart of the transmissions between cities is shown in Figure \ref{fig:0l}.

According to four-tier public health emergency response in mainland China,
and similarly as statement in \cite{ZFC}, we define $\omega(x,y,t)$ as the following step functions to investigate the impact of the population  mobility  between cities on the spatial spreading of COVID-19
\begin{equation}
\omega(x,y,t)=
\left\{
\begin{array}{lll}
\omega(x,y), & T_0\leq t\leq T_1\, \textrm{(no response)},\\
(1-0.95\mu_2)\omega(x,y), &T_1<t\leq T_2\, \textrm{(level I response)},\\[6pt]
(1-0.95\mu_3)\omega(x,y), &T_2<t\leq T_3\, \textrm{(level II response)},\\[6pt]
(1-0.95\mu_4)\omega(x,y), &T_3<t\leq T_4\, \textrm{(level III response)},\\[6pt]
(1-0.95\mu_5)\omega(x,y), & t>T_4\, \textrm{(level IV response)},
\end{array} \right.
\label{a23}
\end{equation}
where we assume that there are five different stages according policy choices, that is, there is no lockdown strategy at
the initial outbreak $[T_0, T_1]$  with normal population mobility $\omega(x, y)$;
Level I response is launched and strict lockdown strategy is carried out during the second stage $(T_1, T_2]$ with bigger effect rate $\mu_2\in [0, 1]$;
In the third stage $(T_2, T_3]$, the emergency response is lowered to level II and  part of transportation restore to restart economic activities and necessary travels.
In the forth stage $(T_3, T_4]$, level III emergency response is carried out, economic activities further restart and travels are permit subject to half of the normal flow.
In the fifth stage $t>T_4$, most of transportation restore but there are still some travel restrictions.

The parameters $\mu_i \in [0, 1]$ for $i=2, 3, 4, 5$  represent weighted averages of restriction levels of transportation between city $x$ and $y$,
where $\mu_i=0$ means usual transportation without any restriction and $\mu_i=1$ means very little transportation permit.
The factor $0.95$ is imposed to reflect the fact that complete lockdown of every city is impossible and impractical,
there are still some necessary business transactions and population movements between cities, see more details in \cite{ZFC}.

Another important control policy is keeping social distancing, which has been proved to be effective to reduce the contact transmission rate and depends on the policy choices, too. In the initial outbreak stage $[T_0, T_1]$, people has no self-protection consciousness, the contact transmission rate $\beta_0$ in this stage is comparatively high; In the second stage, people are required to stay at home and keep social distancing, and the contact transmission rate $\beta(t)$ becomes smaller than that in the first stage, which depends on the intensity and scope of the implementation of prevention and control measures, as well as the severity of the disease and individual's compliance with policies.
In the third stage, part economic activities are reopen and the contact transmission rate may be increasing comparatively.
In the forth stage, most economic activities are reopen, but large scale gathering activities are forbidden,
and in the fifth stage $t>T_4$, mandatory requirements have been canceled but people may be used to take self-protection measures and continue to keep social distancing to some extent. According the staggered properties, we define
\begin{equation}
\beta^*(t)=k\beta(t),\quad \beta (t)=
\left\{
\begin{array}{lll}
\beta_1=\beta_0, & T_0\leq t\leq T_1,\\
\beta_2=(1-0.95\vartheta_2)\beta_0, &T_1<t\leq T_2,\\[6pt]
\beta_3=(1-0.95\vartheta_3)\beta_0, &T_2<t\leq T_3,\\[6pt]
\beta_4=(1-0.95\vartheta_4)\beta_0, &T_3<t\leq T_4,\\[6pt]
\beta_5=(1-0.95\vartheta_5)\beta_0, & t>T_4.
\end{array} \right.
\label{a26}
\end{equation}
Similarly, the parameters $\vartheta_i \in [0, 1]$ for $i=2, 3, 4, 5$  represent weighted averages of restriction levels of social distancing in all cities, and the factor $0.95$ is imposed to reflect the fact that it is impossible to keep social distancing completely even under the strictest segregation policy.

On the other hand, the time from the confirmed cases to be quarantined become shortened,  and then $\gamma_I$ and $\gamma_A$ turn to be bigger over time. Assume that $\gamma_I(t):=\gamma_{I}(0)\gamma (t)=0.1 \gamma (t)$ and $\gamma_A(t):=\gamma_{A}(0)\gamma (t)=0.25 \gamma (t)$, where $\gamma(t)$ gets its minimum value  $1$ (valued at the initial time) and its maximum is $\alpha $ with $\alpha>1$, and then it can be described by the following step functions:
\begin{equation}
\gamma (t)=\left\{
\begin{array}{lll}
\gamma_1=1, & T_0\leq t\leq T_1,\\
\gamma_2=(T_2-T_1)^{-1}\int^{T_2}_{T_1} g(t)dt, &T_1<t\leq T_2,\\[6pt]
\gamma_3=(T_3-T_2)^{-1}\int^{T_3}_{T_2} g(t)dt, &T_2<t\leq T_3,\\[6pt]
\gamma_4=(T_4-T_3)^{-1}\int^{T_4}_{T_3} g(t)dt, &T_3<t\leq T_4,\\[6pt]
\gamma_5=\alpha, & t>T_4,
\end{array} \right.
\label{a27}
\end{equation}
where  $g(t):=\frac \alpha{1+(\alpha-1)e^{-\alpha t}}$ satisfying Logistic type growth.

\section{Basic reproduction number and risk assessment}

We assume that the initial condition
$$(A1)\ S(x,0), E(x,0), A(x,0), I(x,0), R(x,0)\geq 0,\,\sum\limits_{x\in V}E(x,0)+A(x,0)+I(x,0)>0, \,\textrm{and}$$
$$ N(x,0)=N^*(x)>0,\, \textrm{for}\, x\in \Omega,$$
which indicates that the total population of city $x$ at the initial time is $N^*(x)$.
It follows from the qualitative theory of ordinary differential equations \cite{BN89} that a unique nonnegative solution of
\eqref{A3} exists for all $t>0$. The existence and uniqueness of the solution can be also obtained by upper and lower solution methods, see \cite{TLR} and references therein.

As the threshold which reflects the transmission mechanism of the disease, the basic reproduction number $R_0$ is defined as the number of new infections generated by
one infected individual during the entire infectious period in a fully susceptible population \cite{DWR},
and it reflects the ability of an infection spreading during
the early stage of an outbreak.

We first assume that all parameters in $(A1)$ are all independent of $t$.
Recalling that the numbers of population in different cities are different, however the number of population in city $x$ moving to city $y$
is almost equal to that of the population in city $y$ moving to city $x$, so here we assume that $\omega (y, x) N^*(x)=\omega (x, y) N^*(y)$ for $x, y\in V$.

 It is easy to check that system \eqref{A3} subject to assumption $(A1)$ possess a unique disease-free equilibrium
$(N^*(x), 0, 0, 0, 0)$. Since the infected individuals are in the compartments $E, A$ and $I$, therefore, by means of the next
generation matrix approach \cite{DWR},  we can deduce
the vector of new infections $\mathcal{F}$ and the vector of transitions  between compartments $\mathcal{V}$ as follows
\begin{displaymath}
\begin{array}{l}
 \mathcal{F}=\left(
\begin{array}{cc}
 \beta S(I+\xi A)/N+\beta^* \nabla_\omega (\varepsilon I+\xi A) \\
 0\\
 0
 \end{array} \right),
~\mathcal{V}=\left(
\begin{array}{cc}
 -\triangle_\omega E+\sigma E\\
     -\triangle_\omega A+\gamma_A A-(1-\theta)\sigma E \\

    -\varepsilon \triangle_\omega I+ \gamma_I I-\theta\sigma E

 \end{array} \right),
 \end{array}
\end{displaymath}
and the Jacobian matrices of $\mathcal{F}$ and $\mathcal{V}$ are, respectively
\begin{displaymath}
\begin{array}{lll}
~ F=\left(
\begin{array}{ccc}
 0 &  F_{12} & F_{13} \\
0 &   0& 0 \\
0& 0 & 0
\end{array} \right)
=\left(
\begin{array}{ccc}
 0 &  \beta \xi D_n +\beta^* \xi M  & \beta D_n+\beta^*\varepsilon M\\
0 &   0& 0 \\
0 & 0 & 0
\end{array} \right),
\end{array}
\end{displaymath}
and
\begin{displaymath}
\begin{array}{lll}
~ V=\left(
\begin{array}{ccc}
 V_{11} &  0  & 0 \\
-V_{21} &  V_{22} & 0 \\
-V_{31}& 0 & V_{33}
\end{array} \right),
\end{array}
\end{displaymath}
where $D_n$ represents the $n\times n$ identity matrix, the adjacency matrix of the graph $G$: $M=\{\omega_{ij}=\omega(i,j)\}\,(i,j\in V)$  satisfying $\omega_{ii}=0\,(i\in V)$, and
\begin{displaymath}
\begin{array}{lll}
&V_{11}= \textrm{diag} (\sigma+\sum\limits_{j=1}\limits^{n} \omega_{ji})-M,  \\
&V_{22}=\textrm{diag} (\gamma_A+\sum\limits_{j=1}\limits^{n} \omega_{ji})-M, \\
&V_{33}=\textrm{diag} (\gamma_I+\varepsilon \sum\limits_{j=1}\limits^{n} \omega_{ji})-\varepsilon M,\\
&V_{21}=(1-\theta)\sigma D_n,\\
&V_{31}=\theta\sigma D_n,
\end{array}
\end{displaymath}
 Noticing that if $\omega_{ij}\ll 1$, we can easily prove that $V_{11}, V_{22}$ and $V_{33}$ are all irreducible nonsingular $M$-matrices with positive column sums, therefore,  $V_{ii}^{-1}>0$ for $i=1, 2, 3$.
The basic reproduction number for \eqref{A3} is the spectral radius of the next generation matrix $F V^{-1}$, that is, $R_0=\rho (F V^{-1})$, where
\begin{displaymath}
\begin{array}{lll}
~ FV^{-1}=\left(
\begin{array}{ccc}
 0 &  F_{12} & F_{13} \\
0 &   0& 0 \\
0& 0 & 0
\end{array} \right)
\ \left(
\begin{array}{ccc}
 V_{11}^{-1} &  0  & 0 \\
 V_{22}^{-1} V_{21} V_{11}^{-1} &   V_{22}^{-1}& 0 \\
V_{33}^{-1} V_{31} V_{11}^{-1} & 0 & V_{33}^{-1}
\end{array} \right),
\end{array}
\end{displaymath}
therefore,
{\small \begin{equation}
\begin{array}{lll}
R_0&=&\rho(FV^{-1})\\
&=&\rho (F_{12} V_{22}^{-1} V_{21} V_{11}^{-1}+ F_{13}V_{33}^{-1} V_{31} V_{11}^{-1} )\\
&=&\beta \sigma\, \rho (\xi(1-\theta) V_{22}^{-1}  V_{11}^{-1}+\theta V_{33}^{-1} V_{11}^{-1} )+\beta^* \sigma\, \rho (\xi(1-\theta) V_{22}^{-1} M V_{11}^{-1}+\theta \varepsilon V_{33}^{-1}M V_{11}^{-1} )
\end{array}
\label{a261}
\end{equation}}
where the first term accounts for the risk of infection within each city and the second term accounts for
the risk of infection induced by the individual's mobility between cities.

Specially, neglecting the dispersal between cities ($\omega_{ij}=0$), the basic reproduction number with control measures can be expressed as follows
$$R_0 (\omega_{ij}=0) \, =\beta \, \big(\frac{\xi(1-\theta)}{\gamma_A}+\frac{\theta}{\gamma_I}\big).$$

The following dynamical result of problem \eqref{A3} without control measures is obvious, we omit the proof since it is a modification of
Theorem 3.1 in \cite{HDW}.
\begin{thm} Assume that the parameters $\omega, \beta, \gamma_A, \gamma_I$ are all independent of time $t$,\\ $\omega (y, x) N^*(x)=\omega (x, y) N^*(y)$ for $x, y\in V$, and $\omega_{ij}\ll 1$ or the matrix $[\omega_{ij}]$ is irreducible.
If $R_0<1$, the disease-free equilibrium of \eqref{A3} with the initial condition $(A1)$ is globally asymptotically stable.
\end{thm}

However, during the spread of COVID-19 pandemic, consequent policy decisions (such as quarantine, isolation, or
traffic restrictions) have been made according to different situations, the basic reproduction number is no longer a constant and it is time-varying.
 Therefore, it is usually expressed by the effective reproduction number $R_e(t)$ or the basic reproduction number with control measures $R_c(t)$.
 However, owing to characteristics of stages, we have the following staggered reproduction number:

\begin{displaymath}
\begin{array}{lll}
R_e(t)&=&\beta(t) \sigma\, \rho (\xi(1-\theta) V_{22}^{-1}  V_{11}^{-1}+\theta V_{33}^{-1} V_{11}^{-1} )(t)\\
& &+\beta^*(t) \sigma\, \rho (\xi(1-\theta) V_{22}^{-1}M V_{11}^{-1}+\theta \varepsilon V_{33}^{-1}M V_{11}^{-1} )(t)\\
& :=& R_{intra}(t)+R_{inter}(t).
\end{array}
\end{displaymath}

Considering $\omega_{ij}\ll 1$ yields that
$$R_{intra}(t)\approx \beta(t)\,  (\frac {\xi(1-\theta)}{\gamma _A(t)}+ \frac {\theta}{\gamma _I (t)}):=\tilde R_{intra}(t),$$
$$R_{inter}(t)\approx \beta^*(t)\,  (\frac {\xi(1-\theta)}{\gamma _A(t)}+ \frac {\varepsilon \theta}{\gamma _I (t)})\rho (M)(t):=\tilde R_{inter}(t),$$
we then have
$$R_e(t)\approx \tilde R_{intra}(t)+\tilde R_{inter}(t):=\tilde R_e(t),$$
\begin{equation}
\tilde R_{intra}(t)=
\left\{
\begin{array}{lll}
\tilde R_{intra-1}=\beta_0 (\frac {\xi(1-\theta)}{\gamma _A(0)}+ \frac {\theta}{\gamma _I(0)}), & T_0\leq t\leq T_1,\\
\tilde R_{intra-2}=(1-0.95\vartheta_2)\frac{(T_2-T_1)}{\int^{T_2}_{T_1} g(t)dt} \tilde R_{intra-1}, &T_1<t\leq T_2,\\[6pt]
\tilde R_{intra-3}=(1-0.95\vartheta_3)\frac{(T_3-T_2)}{\int^{T_3}_{T_2} g(t)dt} \tilde R_{intra-1}, &T_2<t\leq T_3,\\[6pt]
\tilde R_{intra-4}=(1-0.95\vartheta_4)\frac{(T_4-T_3)}{\int^{T_4}_{T_3} g(t)dt} \tilde R_{intra-1}, &T_3<t\leq T_4,\\[6pt]
\tilde R_{intra-5}=(1-0.95\vartheta_5)\frac{1}{\alpha} \tilde R_{intra-1}, & t>T_4.
\end{array} \right.
\label{a263}
\end{equation}
and
\begin{equation}
\tilde R_{inter}(t)=
\left\{
\begin{array}{lll}
\tilde R_{inter-1}=k\beta_0 (\frac {\xi(1-\theta)}{\gamma_A(0)}+ \frac {\varepsilon\theta}{\gamma_I(0)})\rho (M(T_0)), & T_0\leq t\leq T_1,\\
\tilde R_{inter-2}=(1-0.95\vartheta_2)(1-0.95\mu_2)\frac{(T_2-T_1)}{\int^{T_2}_{T_1} g(t)dt} \tilde R_{inter-1}, &T_1<t\leq T_2,\\[6pt]
\tilde R_{inter-3}=(1-0.95\vartheta_3)(1-0.95\mu_3)\frac{(T_3-T_2)}{\int^{T_3}_{T_2} g(t)dt} \tilde R_{inter-1}, &T_2<t\leq T_3,\\[6pt]
\tilde R_{inter-4}=(1-0.95\vartheta_4)(1-0.95\mu_4)\frac{(T_4-T_3)}{\int^{T_4}_{T_3} g(t)dt} \tilde R_{inter-1}, &T_3<t\leq T_4,\\[6pt]
\tilde R_{inter-5}=(1-0.95\vartheta_5)(1-0.95\mu_5)\frac{1}{\alpha} \tilde R_{inter-1}, & t>T_4.
\end{array} \right.
\label{a264}
\end{equation}
It is obvious from \eqref{a263} and \eqref{a264} that staged policy choices for lockdown and social distancing both lead to the change of the basic reproduction number,
which reveals that appropriate control measures can reduce the risk of pandemic and slow the spread of the COVID-19.

\section{Robustness of the simulation results}

We retrieved daily numbers of COVID-19 cases in Wuhan and 18 other cities/regions in mainland China that reported more cases than other cities/regions by January 23, 2020 (Wuhan, Chongqing, Beijing, Xiaogan, Shanghai, Taizhou, Shenzhen, Huanggang, Jingzhou, Jingmen, Changsha, Zhuhai, Guangzhou, Chengdu, Foshan, Beihai, Hangzhou, Hefei, Wenzhou). We obtained the population flow data from Baidu platform (https://qianxi.baidu.com).

\begin{figure}
    \centering
    \includegraphics[width=1\textwidth]{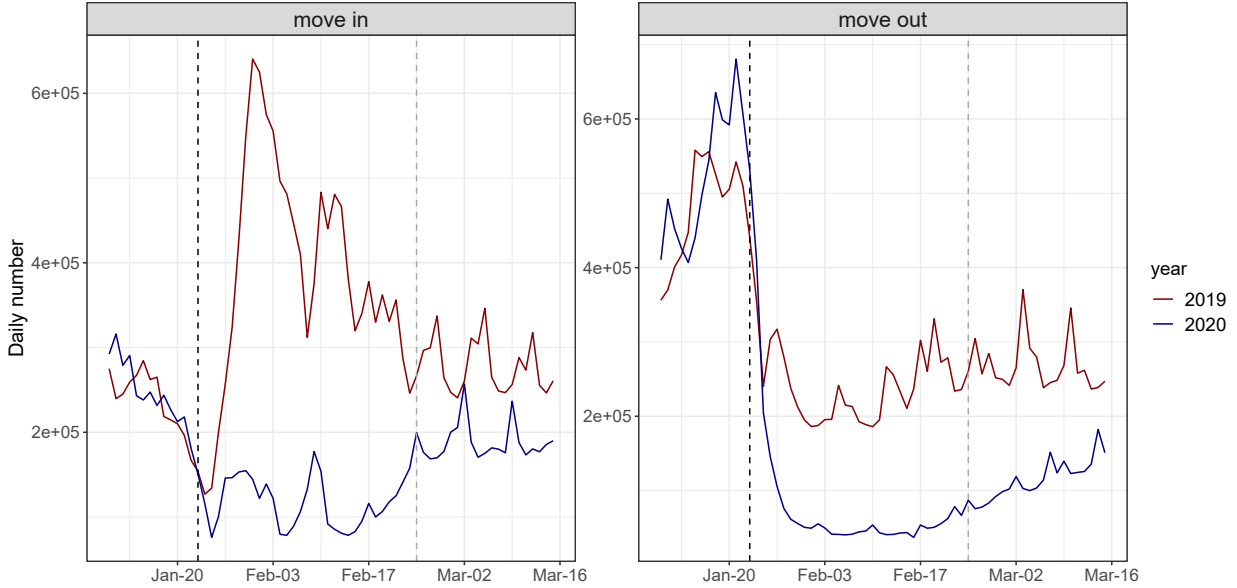}
    \caption{Comparison of the number of population flow between 2019 and 2020 in a typical city (Guangzhou), during January 10 to March 15, 2020 and the same period in 2019 (according to the lunar calendar). The back dash lines show the implementation of level I response and gray lines represent the begin of level II response. }
    \label{fig:my_label1}
\end{figure}

\begin{figure}
    \centering
    \includegraphics[width=1\textwidth]{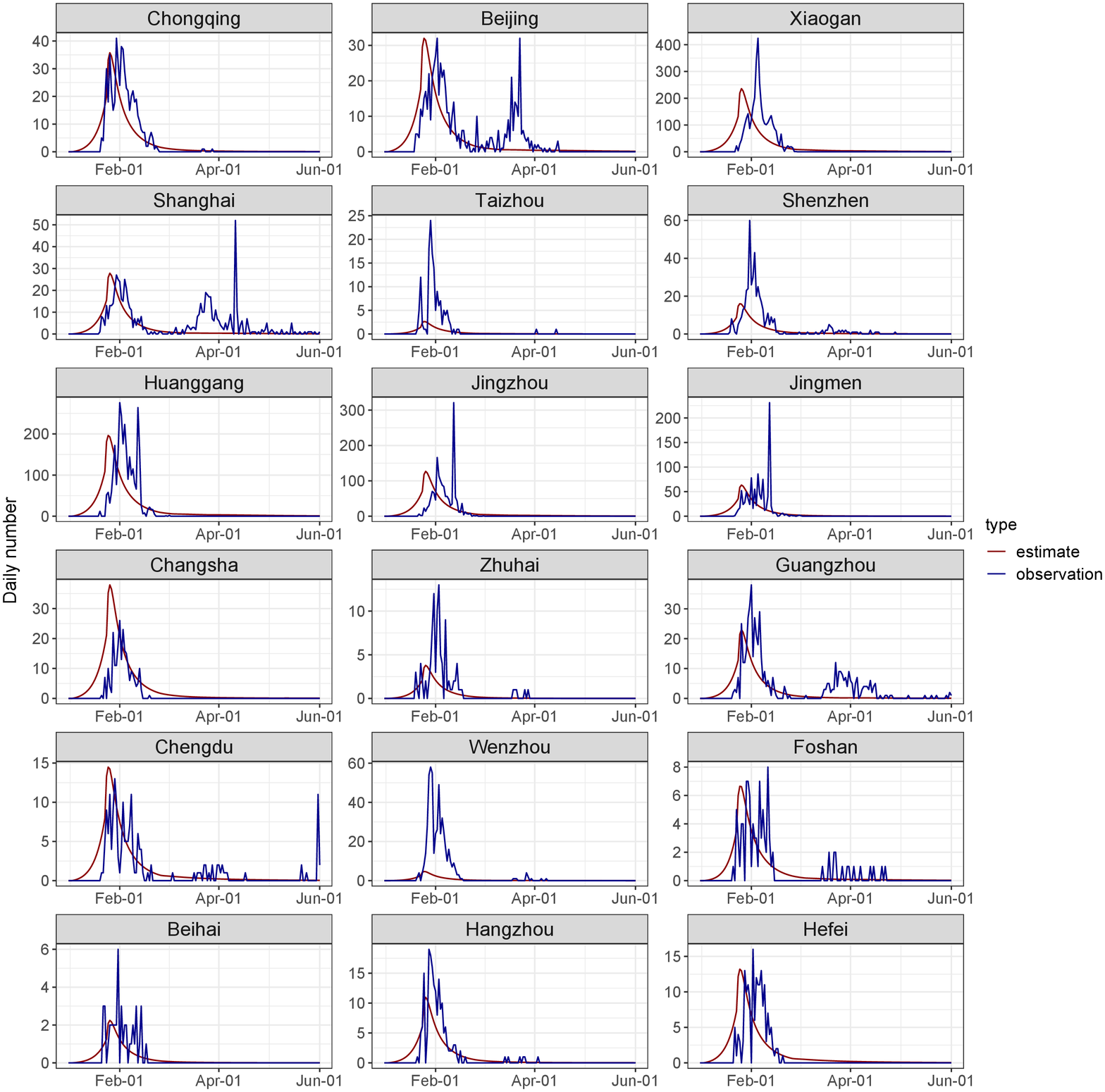}
    \caption{Plots of model estimates and observations for the daily numbers of COVID-19 cases in 18 cities/regions, during January 1, 2020 to June 1, 2020. Each panel contains the actual reported numbers (blue line), and the fitted line (red line).}
    \label{fig:my_label2}
\end{figure}

Figure \ref{fig:my_label1} shows a comparison of the number of population flow between 2019 and 2020 in a typical city (Guangzhou). We can clearly find that after the implementation of the level I response, the number of people move in and move out from Guangzhou in 2020 was significantly lower than that in 2019. The gap was slightly narrowed when the level II response is launched. Daily number of confirmed COVID-19 cases in each city was assumed to follow a Poisson distribution. We calculate the maximum likelihood estimates of $\mu_i,\vartheta_i,\alpha,\varepsilon$ and $k$ by fitting model to the number of newly reported cases before June 1, 2020. Then we obtained the 95\% confidence intervals (95\% CI) of these parameters by applying the profile likelihood estimation framework with a cutoff threshold provided by Chi-square quantile. Please note that reported cases in Wuhan was not included in the fitting process since researches \cite{LPC, ZhaoS} suggested that the under-reporting was likely to occur in the early stage of outbreak in Wuhan and confirmed cases increased dramatically after February 12, 2020 in Wuhan, due to the changing of the diagnosis criteria. The fitting results are largely consistent with daily numbers of newly reported cases by June 1, 2020 in cities/regions selected (Figure \ref{fig:my_label2}).

\begin{figure}
    \centering
    \includegraphics[width=1\textwidth]{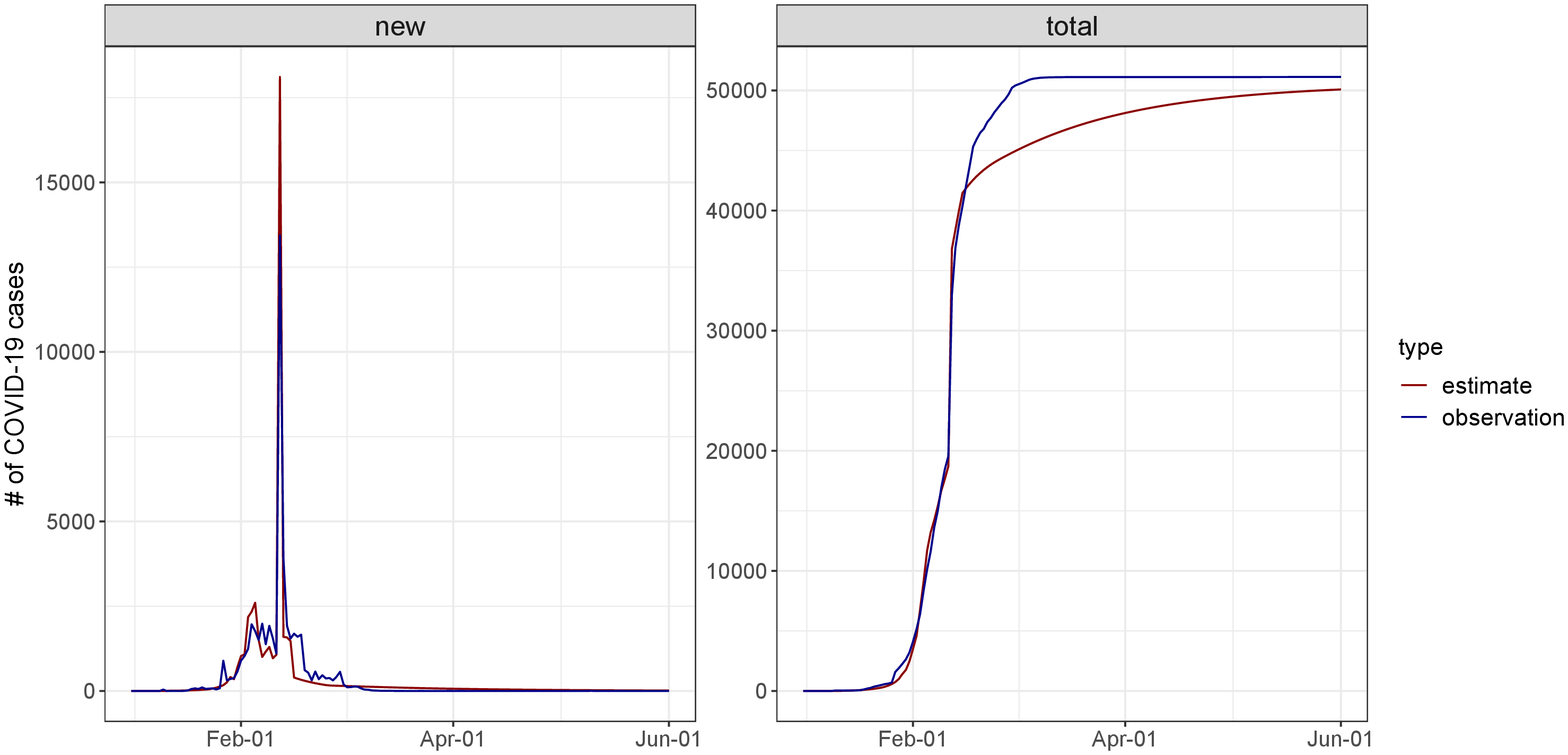}
    \caption{Plots of model estimates and observations for the daily numbers of COVID-19 cases in Wuhan, during January 1, 2020 to June 1, 2020. Each panel contains the actual reported numbers (blue line), and the fitted line (red line).}
    \label{fig:my_label3}
\end{figure}

As for Wuhan, we assumed a ascertainment rate of 1.8\% \cite{WLB} on January 1 and increased
linearly to 14\% \cite{LPC} on January 23 and then increased linearly to 80\% before February 12
when diagnosis criteria was changed in Wuhan. We assumed a 14-day reporting lag of
COVID-19 cases before 17 January \cite{ZZW} and increased linearly to 3 days before February 12.
Due to the diagnosis criteria change, we assumed that cases undocumented in the previous
20 days were reported on February 12. In addition, ascertainment rate and reporting lag
after February 12 was assumed to be 100\% and 0 days, respectively. Figure \ref{fig:my_label3}  presents that the adjusted estimates of daily reported and cumulative number of infections was close to
the observations. Table 2 lists the value and 95\% CI of estimated parameters.

{\small\begin{table}[htbp]

\small
\centering
\begin{threeparttable}
\caption{\label{table2}Maximum likelihood estimates and 95$\%$ CI of parameters estimated from the model. }
\setlength{\tabcolsep}{12mm}
\begin{tabular}{|c|c|c|c|}
\hline
Parameters & Estimate  & 95$\%$ CI      \\ \hline

$\mu_2,\vartheta_2$ & 0.81 & (0.80,0.82) \\  \hline
$\mu_3,\vartheta_3$ & 0.68 & (0.62,0.7) \\  \hline
$\mu_4,\vartheta_4$ & 0.59 & (0.55,0.60) \\  \hline
$\mu_5,\vartheta_5$ & 0.15\tnote{1} & / \\  \hline
$\alpha$ & 1.42 & (1.41,1.43) \\  \hline
$\varepsilon$ & 0.04 & (0.03,0.16) \\  \hline
$k$ & 72 & (66,78) \\  \hline

\end{tabular}
\begin{tablenotes}
\footnotesize
\item[1] Since none of the selected cities has entered the Level IV response yet, We assumed the value of $\mu_5$ and $\vartheta_5$ according to the content of the policy.
\end{tablenotes}
\end{threeparttable}
\end{table}}

\begin{figure}
    \centering
    \includegraphics[width=1\textwidth]{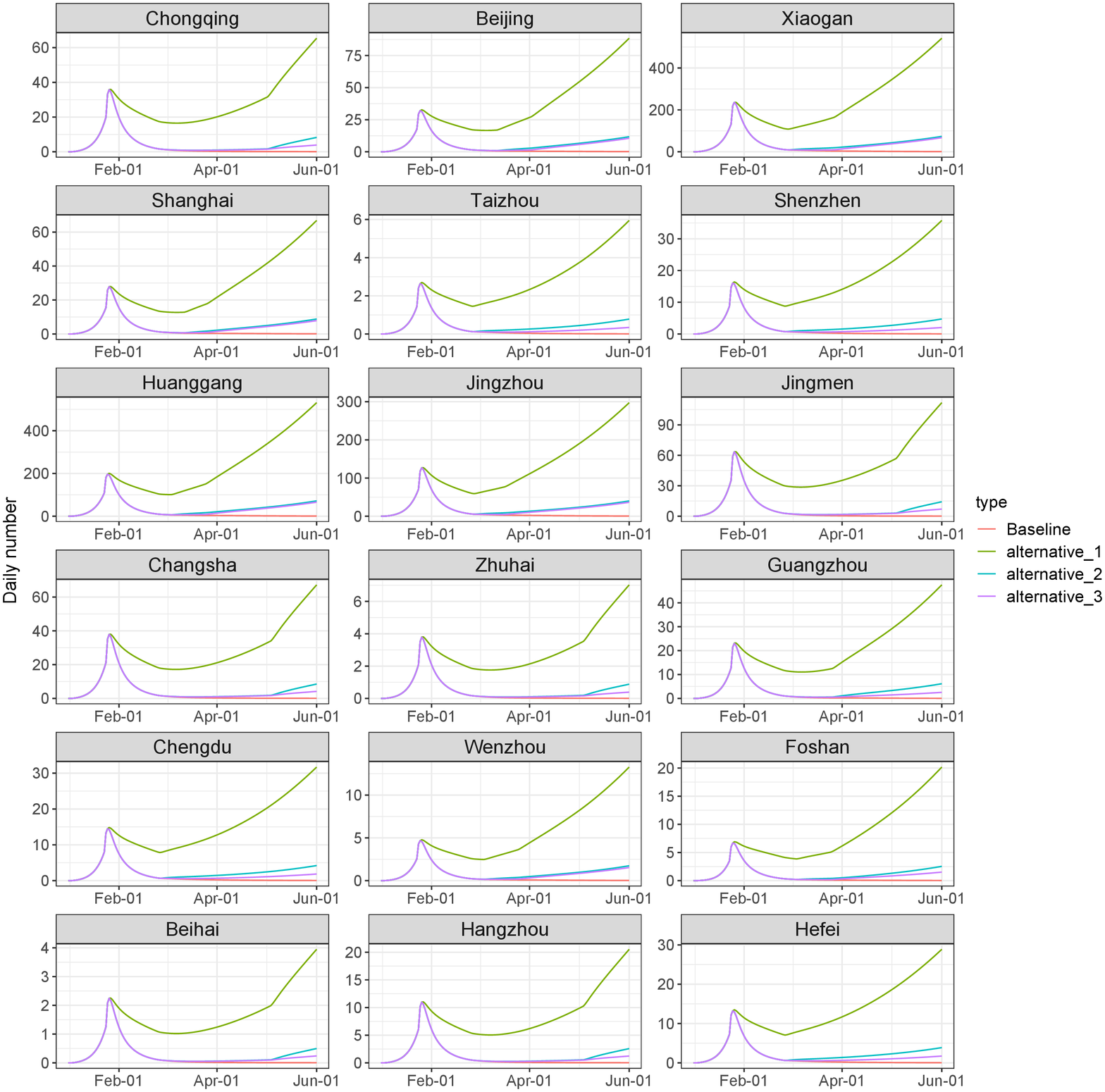}
    \caption{Evaluation of lockdown and social distancing policy under different scenarios in 18 cities/regions by June 1, 2020. We considered the current policy (pink), together with three other scenarios: The overall response intensity decreased to 70\% (green), all the response intensity decreased to 70\% after Level I response (light blue), and all the response intensity decreased to 70\% after Level II response (purple).}
    \label{fig:my_label4}
\end{figure}

\begin{figure}
    \centering
    \includegraphics[width=1\textwidth]{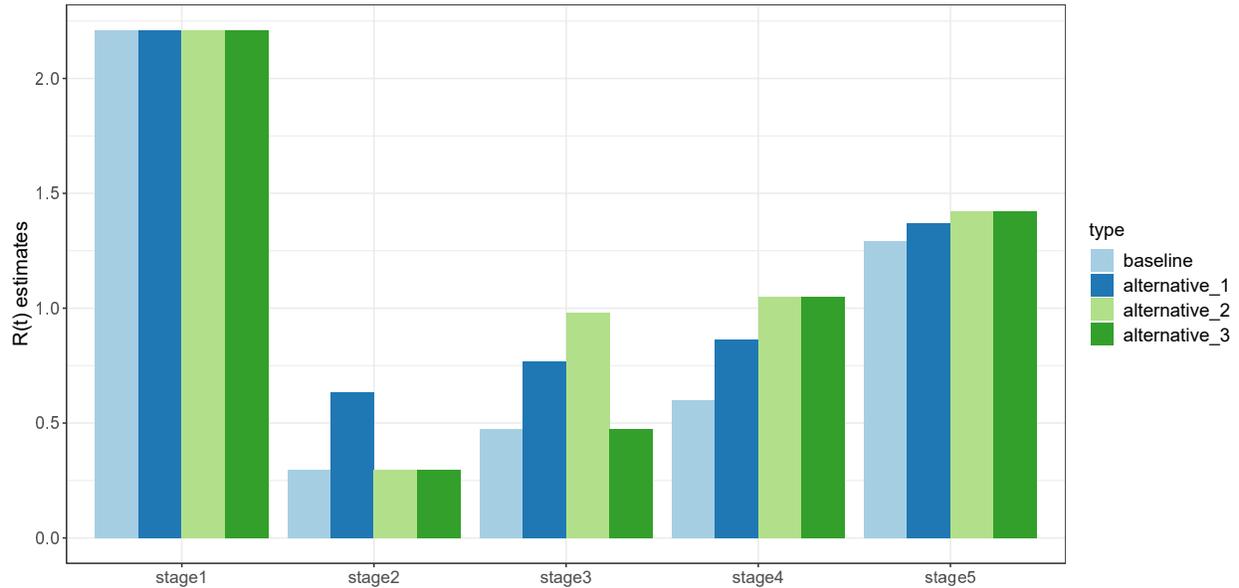}
    \caption{ Estimates of $R_e(t)$ for each stage under different scenarios: The baseline scenario (light blue bar); The overall response intensity decreased to 70\% (dark blue bar); All the response intensity decreased to 70\% after Level I response (light green bar); and all the response intensity decreased to 70\% after Level II response (dark green bar).}
    \label{fig:my_label5}
\end{figure}

 As shown in Table 3, we simulated three additional scenarios other than baseline condition: The overall response intensity decreased to 70\% (alternative\_1), all the response intensity decreased to 70\% after Level I response (alternative\_2) and all the response intensity decreased to 70\% after Level II response (alternative\_3). In Figure \ref{fig:my_label4}, our simulations implies that if intensity of lockdown and  social distancing restrictions were 70$\%$ of the current situation, the duration of the epidemic would be significantly prolonged. A low intensity of the response is likely to lead to the second wave of the epidemic. We also estimated the basic reproduction numbers $R_e(t)$ in five stages under these scenarios in Figure \ref{fig:my_label5}, which suggests that stringent lockdown and social distancing policy provide a smaller basic reproduction number,
reduce the infection risk and low the virus spreading speed.

{\small\begin{table}[htbp]
\small
\centering
\caption{\label{table3}Intensity of response and restriction of social distancing. }
\setlength{\tabcolsep}{7mm}
\begin{tabular}{|c|c|c|c|c|}
\hline
Scenarios & $\mu_2,\vartheta_2$  & $\mu_3, \vartheta_3$ & $\mu_4, \vartheta_4$ & $\mu_5, \vartheta_5$     \\ \hline

Baseline & 0.81 & 0.68 & 0.59 & 0.15 \\  \hline
alternative\_1 & 0.57 & 0.48 & 0.41 & 0.11 \\  \hline
alternative\_2 & 0.81 & 0.48 & 0.41 & 0.11 \\  \hline
alternative\_3 & 0.81 & 0.68 & 0.41 & 0.11 \\  \hline

\end{tabular}
\end{table}}

\begin{figure}
    \centering
    \includegraphics[width=1\textwidth]{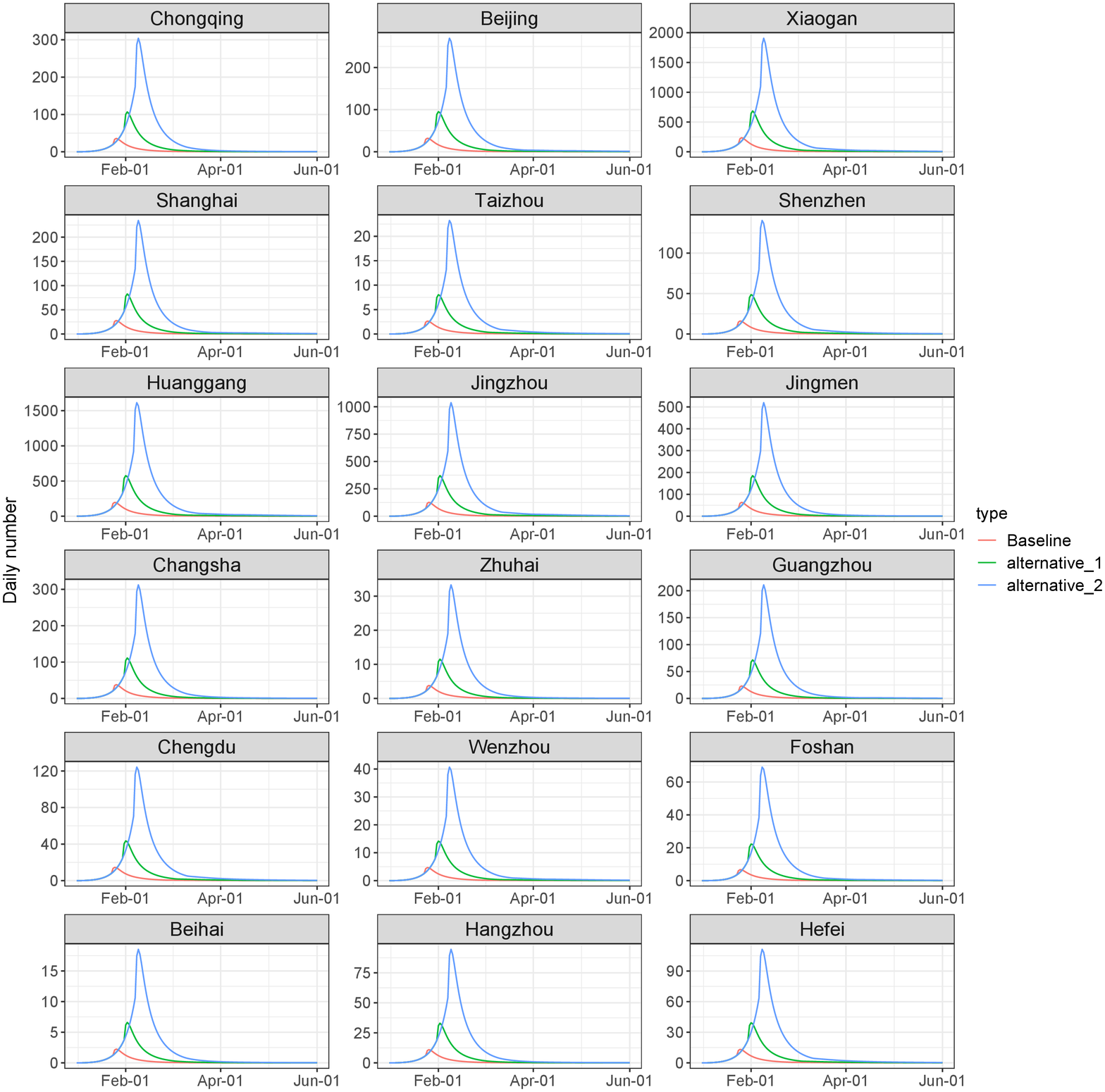}
    \caption{Evaluation of lockdown and social distancing policy under different scenarios in 18 cities/regions by June 1, 2020. We considered the current policy (red), together with two other scenarios: a 7-day delay of the intervention (green) and a 14-day delay of the interventions (blue).}
    \label{fig:my_label6}
\end{figure}

To examine the impact of the timing of the emergency response, Figure \ref{fig:my_label6} presents the simulated results for additional scenarios that all responses were deferred for one week (7 days) and two weeks (14 days). It shows that a later implementation of the lockdown and social distancing policy significantly increased the total number of infections. Compared with the baseline scenario, a 7-day and a 14-day delay of the interventions may lead to a three times and ten times of the peak daily infections, and postpone the peak of daily infections by around 7 days and 14 days, respectively.

\section{Discussion}

The modeling and simulations reveal that Level I public health emergency response was essential and proven to be a successful strategy
for high-risk cities. Level II response is suitable for medium-risk regions where the basic reproduction number has dropped
significantly, and the social and economical needs to restart, but the basic reproduction number is still near $1$. It means that the ongoing COVID-19 pandemic has not been completely contained, prevention and control measures cannot be relaxed, and the self-protection measures is still needed to prevent the second wave of the epidemic.
Level III public health response is effective for lower-risk regions, where there has been no new confirmed case for six consecutive days, and most of the patients have been discharged from hospitals.
Level IV will be the normalize life for several months or even longer until the vaccine has been successfully developed and put into use. These research results show that it is necessary to adopt different levels of public health emergency response to prevent and control the epidemic in multi-stage and multi-region. The success of prevention and control of the COVID-19 in Wuhan and other provinces of mainland China also reveals that this is an effective approach to inhibit the large-scale and rapid spread of COVID-19 pandemic.

 Mathematically, the stricter the control strategies, the smaller the staggered reproduction number. However, the implementation of the policy is also a double-edged sword. The strict control strategies will certainly lead to heavy economic losses and great inconvenience to people's work and life.
 It was easimated in \cite{you} that the total monthly economic losses during the first month of the lockdown of Wuhan from January 23 to
February 23, 2020 reach 177.0413 billion yuan. Also, it was also shown in \cite{yang} that the lockdown also increased the mental stress of residents.
Strict traffic restrictions and suspension of non emergency medical services are likely to delay hospitalization of non-COVID-19 patients, some of whom may already be in critical situation \cite{hui}. Therefore, adopting a multi-stage control strategies according to the public health emergency response levels is not only conducive to the control of the epidemic, but also conducive to social development and the normalization of people's lifestyle.
How to balance epidemic control, economic activities and social development is a difficult choice for every decision maker. We will discuss it in our coming work.


\begin{thebibliography}{99}%\setlength{\itemsep}{-1ex}
\bibitem{ACW1}
D. Acemoglu, V. Chernozhukov, I. Werning, et al., Optimal targeted lockdowns in a multi-group SIR model, NBER Working Paper No, 2020, 27102.

\bibitem{ACW2}
D. Acemoglu, V. Chernozhukov, I. Werning, et al., A multi-risk SIR model with optimally targeted lockdown, National Bureau of Economic Research, 2020.

\bibitem{BN89}
F. Brauer,  J. A. Nohel, The Qualitative Theory of Ordinary Differential Equations, Dover Publications, 1989.

\bibitem{BHR}
P. Block, M. Hoffman, I. J. Raabe, et al., Social network-based distancing strategies to flatten the COVID-19 curve in a post-lockdown world, Nature Human Behaviour, 2020, 1-9.

\bibitem{BHL}
F. Bauer, P. Horn, Y. Lin, G. Lippner, D. Mangoubi, S. T. Yau, Li-Yau inequality on graphs, J.
Differential Geom., 99 (2015), 359-405.


\bibitem{BD}
B. Buonomo, R. Della Marca, Modelling information-dependent social behaviors in response to lockdowns: the case of COVID-19 epidemic in Italy, medRxiv, 2020.

\bibitem{EMPE}
S. Eikenberry, M. Mancuso, E. Iboi, T. Phan, K. Eikenberry, Y. Kuang, E. Kostelich, and A. Gumel, To mask or not to mask: Modeling the potential for face mask use by the general public to curtail the COVID-19 pandemic, Infectious Disease Modelling, 5(2020), 293-308.


\bibitem{EEL}
 S. Eubank, I. Eckstrand, B. Lewis, et al., Impact of Non-pharmaceutical Interventions (NPIs) to Reduce COVID-19 Mortality and Healthcare Demand, Bull. Math. Biol.,  82 (2020), Paper No. 52, 7 pp.

\bibitem{FNP}
Y. Q. Fang, Y. T. Nie, M. Penny, Transmission dynamics of the COVID-19 outbreak and effectiveness of government interventions: A data-driven analysis, Journal of medical virology, 92.6 (2020): 645-659.

\bibitem{FN}
N. Ferguson, et al., Report 9: Impact of non-pharmaceutical interventions (NPIs) to reduce COVID19 mortality and healthcare demand, (2020).

\bibitem{GLIN}
J. Ge, Z. G. Lin, A SEIR networked model incorporating intracity and intercity transmissions, to appear.


\bibitem{GNHL}
W. Guan, Z. Ni, Y. Hu, et al., Clinical characteristics of 2019 novel coronavirus infection in China, medRxiv. 2020:2020.02.06.20020974.

\bibitem{HPS}
S. B. He, Y. X. Peng, K. H. Sun, SEIR modeling of the COVID-19 and its dynamics, Nonlinear Dynamics, (2020): 1-14.


\bibitem{HDW}
Y. H. Hsieh, P. van den Driessche, L. Wang,
Impact of travel between patches for spatial spread of disease, Bull. Math. Biol., 69 (2007), no. 4, 1355-1375.

\bibitem{HWLR}
C. Huang, Y. Wang, X. Li, et al., Clinical features of patients infected with 2019 novel coronavirus in Wuhan, China, The Lancet, 2020.

\bibitem{hui}
D. S. Hui, E. I. Azhar, T. A. Madani, et al., The continuing 2019-nCoV epidemic threat of novel coronaviruses to global health¡ªThe latest 2019 novel coronavirus outbreak in Wuhan, China, International Journal of Infectious Diseases, 91(2020), 264-266.

\bibitem{JS}
S. Jain, A. Sinha, Adwitiya, Identification of influential users on Twitter: A novel weighted correlated influence measure for Covid-19,
Chaos Solitons Fractals,  139  (2020), 110037.

\bibitem{Krae}
M. U. G. Kraemer, C. H. Yang, B. Gutierrez, et al., The effect of human mobility and control measures on the COVID-19 epidemic in China, Science, 2020, 368(6490): 493-497.

\bibitem{LSM}
S. Lalwani, G. J. Sahni, B. Mewara, R. Kumar, Predicting optimal lockdown period with parametric approach using three-phase maturation SIRD model for COVID-19 pandemic, Chaos Solitons Fractals,  138 (2020), 109939.

\bibitem{LKK}
H. Lau, V. Khosrawipour, P. Kocbach, et al., The positive impact of lockdown in Wuhan on containing the COVID-19 outbreak in China, Journal of Travel Medicine, 2020, 27(3): taaa037.


\bibitem{Zhu}
J. Li, P. Yuan, J. Heffernan, et al., Observation wards and control of the transmission of COVID-19 in Wuhan, (Submitted) Bull World Health Organ., E-pub: 9 April. doi: http://dx.doi.org/10.2471/BLT.20.258152.

\bibitem{LS}
M. Li, Z. Shuai, Global-stability problem for coupled systems of differential equations on networks, J. Differential Equations, 248 (2010), 1-20.

\bibitem{LPC}
R. Y. Li, S. Pei, B. Chen, Y. M. Song, T. Zhang,  et al., Substantial undocumented infection facilitates the rapid dissemination of novel coronavirus (SARS-CoV-2), Science, 2020, 368(6490): 489-493.

\bibitem{LPS}
K. Linka, M. Peirlinck, F. Sahli Costabal, et al., Outbreak dynamics of COVID-19 in Europe and the effect of travel restrictions, Computer Methods in Biomechanics and Biomedical Engineering, 2020: 1-8.

\bibitem{NHC}
http://www.nhc.gov.cn/bgt/s9509/200905/9297d07be9a5487e93a7c162fa15892f.shtml

\bibitem{NHC1}
http://www.nhc.gov.cn/xcs/fkdt/202004/83ce73decd71499fa5ac4b69eaa71a6d.shtml

\bibitem{PCV}
R. Pastor-Satorras, C. Castellano, P. Van Mieghem, A. Vespignani, Epidemic processes in complex
networks, Rev. Mod. Phys., 87 (2015), 925-986.

\bibitem{PLC}
M. Peirlinck, K. Linka, F. S. Costabal, et al., Outbreak dynamics of COVID-19 in China and the United States, Biomechanics and modeling in mechanobiology, 2020: 1.

\bibitem{PSZ}
N. Picchiotti, M. Salvioli, E. Zanardini, et al., COVID-19 Italian and Europe epidemic evolution: A SEIR model with lockdown-dependent transmission rate based on Chinese data, Available at SSRN 3562452, 2020.

\bibitem{RAl}
J. Riou, C. L. Althaus, Pattern of early human-to-human transmission of Wuhan 2019 novel coronavirus (2019-nCoV), December 2019 to January 2020. EuroSurveill 2020:25.

\bibitem{RVH}
W. C. Roda, M. B. Varughese, D. Han, M. Y. Li, Why is it difficult to accurately predict the COVID-19 epidemic? Infectious Disease Modelling, (2020).

\bibitem{songlou}
P. F. Song, Y. Lou, L. P. Zhu, et al,
Multi-stage and multi-scale patch model and the case study of novel coronavirus, Acta Mathematicae Applicatee Sinica, 43, 2020, 174-199. in Chinese.


\bibitem{Sunw}
G. Q. Sun, S. F. Wang, M. T. Li, et al., Transmission dynamics of COVID-19 in Wuhan, China: effects of lockdown and medical resources, Nonlinear Dyn., 2020,
https://doi.org/10.1007/s11071-020-05770-9.

\bibitem{TLR}
C. R.  Tian, Z. H. Liu, S. G. Ruan, Dynamical behavior of a weighted networked SEIR epidemic model, to appear.

\bibitem{DWR}
P. van den Driessche and J. Watmough, Reproduction numbers and sub-threshold endemic equilibria
for compartmental models of disease transmission, Math. Biosci., 180 (2002), 29-48.


\bibitem{WHO}
https://www.who.int/emergencies/diseases/novel-coronavirus-2019

\bibitem{WLB}
J. T. Wu, K. Leung, M. Bushman, N. Kishore, R. Niehus, et al., Estimating clinical severity of COVID-19 from the transmission dynamics in Wuhan, China. Nature Medicine, 26(4)(2020), 506-510.

\bibitem{yang}
Y. Yang, W. Li, Q. Zhang, L. Zhang, T. Cheung, Y. T. Xiang, Mental healthservices for older adults in China during the COVID-19 outbreak, The Lancet Psychiatry, 2020; 7(4):e19.

\bibitem{you}
S. You, H. Wang, M. Zhang, et al. Assessment of monthly economic losses in Wuhan under the lockdown against COVID-19. Humanit Soc Sci Commun 7, 52 (2020). https://doi.org/10.1057/s41599-020-00545-4.

\bibitem{Feng}
H. Zhao and Z. L. Feng, Staggered release policies for COVID-19 control:
Costs and benefits of relaxing restrictions by age and risk, Mathematical Biosciences (2020), doi:https://doi.org/10.1016/j.mbs.2020.108405.

\bibitem{ZFC}
H. Zhao, Z. L. Feng, Castillo-Chavez C, et al., Staggered release policies for covid-19 control: Costs and benefits of sequentially relaxing restrictions by age, arXiv preprint arXiv:2005.05549, 2020.

\bibitem{ZJY}
Y. Z. Zhang, B. Jiang, J. M. Yuan, et al., The impact of social distancing and epicenter lockdown on the COVID-19 epidemic in mainland China: A data-driven SEIQR model study, medRxiv, 2020.


\bibitem{ZhaoS}
S. Zhao, S. S. Musa, Q. Y. Lin, et al., Estimating theunreported number of novel coronavirus (2019-nCoV) cases in China in the first half of January 2020: a data-driven modelling analysisof the early outbreak, Journal of clinical medicine, 2020; 9(2): 388.

\bibitem{ZXT}
W. K. Zhou, A. L. Wang, F. Xia, Y. N. Xiao, S. Y. Tang, Effects of media reporting on mitigating spread of COVID-19 in the early phase of the outbreak, Mathematical Biosciences and Engineering, 2020, 17(3): 2693-2707.

\bibitem{ZZW}
N. Zhu, D. Zhang, W. Wang, X. Li, B. Yang, et al., A novel coronavirus from patients with pneumonia in China, 2019, New England Journal of Medicine, 2020.


\bibitem{Zou}
L. Zou, S. G. Ruan, A patch model of COVID-19: the effects of containment on Chongqing, Acta Mathematicae Applicatee Sinica, 2020, 43(02): 310-323. in Chinese.



\end{thebibliography}
\end{document}